\documentstyle[12pt]{article}

\catcode`\@=11
\def\numberbysection{\@addtoreset{equation}{section}
        \def\theequation{\thesection.\arabic{equation}}}
\def\beq{\begin{equation}}
\def\eeq{\end{equation}}
\def\half{\frac{1}{2}}
\numberbysection
\begin{document}
\begin{titlepage}
\begin{center}
\hfill CERN-TH / 95-192 \\
\vskip 1.in
{\Large
\bf (2+1)-Gravity with moving particles in  \\
an instantaneous gauge \footnote{Work supported in part by M.U.R.S.T., Italy.
\\ ${}^{**}$ On sabbatical leave of absence from Dipartimento di Fisica,
Universit\`a di Firenze and INFN, Sezione di Firenze, Italy.\\
{} \\
CERN-TH / 95-192 \\
July 1995 } }
\vskip 0.2in
A. Bellini \\
[.2in] {\em Dipartimento di Fisica, Universit\`a di Firenze, Italy,}
\vskip 0.2in
M. Ciafaloni${}^{**}$
\vskip 0.2in
{\em Theoretical Physics Division, CERN, Geneva}
 \vskip 0.2in  and  \vskip 0.2in
P. Valtancoli \\
\vskip 0.2in {\em Dipartimento di Fisica, Universit\`a di Firenze \\
and INFN, Sezione di Firenze, Italy} \\
\end{center}
\vskip .5in
\begin{abstract}
By defining a regular gauge which is conformal-like and provides instantaneous
field propagation, we investigate classical solutions of (2+1)-Gravity coupled
to arbitrarily moving point-like particles. We show how to separate field
equations from self-consistent motion and we provide a solution for the metric
and the motion in the two-body case with arbitrary speed, up to second order
in the mass parameters.
\end{abstract}
\vfill
\end{titlepage}
\pagenumbering{arabic}
%***************************************************************************
%
%
%***************************************************************************
In the past few years, much attention has been devoted to the gravitational
problem in 2+1-dimensions \cite{a1}-\cite{a9}, mostly because it shows a few
simplifying features which may allow a treatment of the quantum problem.

First of all, the three-dimensional geometry is characterized by the fact
that the space is flat outside the matter sources \cite{a1}-\cite{Gott}
. This implies that the
dynamics of pointlike particles can be made locally trivial and should be
determined at large by the global structure of space-time, as suggested
by its connection with the ( topological ) Chern-Simons Poincar\'e gauge
theory \cite{a3}-\cite{a4}.

Secondly, the perturbative quantum problem \cite{a4}-\cite{a7} is characterized
by the absence of (transverse) gravitons. This lack of graviton radiation makes
the infrared properties of the theory much "potential" like and may allow a
quantum treatment with naive definition of matter asymptotic states.

The first feature has been used to construct general classical solutions for
N moving \hskip 0.5 in
  ( massive ) particles \cite{a8}-\cite{a9}. For a single particle,
one has a cut Minkowskian space-time where the two edges of the cut are related
by a rotation in the static case, corresponding to the deficit angle of a
conical space and, more generally, by a Lorentz transformation for
non-vanishing speed.

Many particles solution can thus be obtained by superimposing in a linear
way the various cuts or tails attached to the particle trajectories. This
simple linear description is obtained at the expense of singularities and/or
multivaluedness of the connection matrix ( and of the metric tensor ) along
the above-mentioned cuts or tails, even for the case of massive particles,
where such singularities are not possibly induced by the $v \rightarrow c $
limit.

In other words, this class of exact solutions is obtained by choosing a
singular gauge in which the metric and the connection are singular even
outside the particle trajectories.

By contrast, N-particle classical solutions are fully non-linear and
non trivial in regular gauges, in which the particle sites are the only
isolated singularities. A method for constructing the non linear
coordinate transformation from singular to regular gauges was given in
\cite{a8}, where also the explicit solution was exhibited in the massless
limit (see also \cite{a10})

The purpose of the present paper is to investigate classical metric and motion
in a regular gauge which reduces to the conformal one in
the static limit, and is similar to the one used by two of us \cite{a11} to
discuss the quasi-static case to all orders.

The characteristic feature of this gauge is that it yields an instantaneous
propagation for arbitrarily moving particles, and also has a diagonal space
part, thus generalizing the conformal gauge. As a consequence, we are able to
split the Einstein equations in a set of four, which determines the metric,
and in a remaining set, which determines the motion.

Needless to say, the \  considerable simplification mentioned \ above is due to
the 3-dimensio\-nal nature of the problem, and in particular to the fact that,
for a given wave vector there is only one transverse direction, which is unable
to propagate physical tensor waves.
Nevertheless, it is interesting that this procedure allows to find - at least
perturbatively - a regular metric, and to set up the equations of motion in a
Newtonian way, for an arbitrary set of moving particles.

In this paper we limit ourselves to the two-body problem with masses
$ m_1 , m_2 $ and arbitrary speed, and we solve for the metric and for the
motion up to second order in the mass parameters $Gm_{i} $. We also provide the
corresponding expression for the scattering angle.

The contents of the paper are as follows. In Sec. II we define our gauge
choice,
and we describe the corresponding field equations with instantaneous
propagation
and the particle's equations of motion.
In Sec. III we set up the perturbative treatment of our problem
and we derive the first order results for both metric and motion. In Sec. IV
we derive our main results for the metric tensor and the
connection for arbitrary speed, and up to second order in $Gm_i$.
Finally, in Sec. V we discuss the equations of motion and the ensuing
scattering angle up to second order and we outline possible
developments.
Some details of the calculations are deferred to Appendices A and B.
%*************************************************************************
%
\section{ An instantaneous gauge for moving particles:
general features}
%
%*************************************************************************

For the purpose of orientation, let us recall the static many-particle
solutions in (2+1)-dimensions. They were first found \cite{a1}-\cite{a2}
in the conformal gauge, defined by
\beq g_{\mu\nu} = \left( \begin{array}{cc} 1 & 0 \\
0 & - e^{2\phi} \delta_{ij} \end{array} \right)
( \mu , \nu = 0, 1, 2; i,j = 1, 2 ) . \label{con} \eeq
In this gauge, for one particle at rest in the origin one simply finds
\beq
e^{2\phi} = \alpha^2 R^{- 8 G m}, \ \ \ \ \alpha = 1 - 4 G m ,
\label{sta} \eeq
where $R^2 = {\bf x}^2$, so that
\beq ds^2 = dt^2 - \alpha^2 R^{- 8 G m} {d {\bf x}}^2  . \label{met} \eeq
This proper-time interval can be related, by a redefinition of the radial
coordinate
\beq r = R^{\alpha} \label{coo} \eeq
to the conical gauge expression
\beq
ds^2 = dt^2 - dr^2 - \alpha^2 r^2 d \theta^2 ,
\label{oni} \eeq
thus yielding the customary description of space-time characterized by the
deficit angle
\beq
2\pi - 2\pi \alpha = 8 \pi G m .
\label{def} \eeq
For many particles at rest at ${\bf x} = {\bf \xi}_i$, the conformal factor is
multiplicative, or equivalently $\phi$ is additive, i.e.,
\beq \phi = - 4G \sum_i m_i \log | {\bf x} - { \bf \xi}_i | + {\rm \ const. } .
\label{fac} \eeq
The corresponding conical description was given in \cite{a8} by the
$\Lambda$-mapping method, and involves a slightly more complicated
coordinate transformation.

Here, we are interested in generalizing the conformal gauge to allow a
reasonably simple description of moving particles. In this general case, we
can always reduce the spatial part of the metric to diagonal form, or, by using
complex $z,\bar z $ coordinates, we can set
\beq  g_{zz} =  g_{\bar z\bar z} = 0 ,
\label{omp} \eeq
and we still have the freedom of an additional gauge condition. However, in
general the mixed space-time components will be non-vanishing, and we
parametrize
\beq  g_{00} = {\alpha}^2 - e^{2\phi}\beta\bar \beta , \ \
g_{0z} = \half\bar \beta e^{2\phi} , \ \ g_{0\bar z} = \half\beta e^{2\phi} ,
\ \ g_{z\bar z} = -\half e^{2\phi}  , \label{par} \eeq
where $\alpha(z,\bar z,t)$ and $ \phi(z,\bar z,t) $ are real functions and
$\beta(z,\bar z,t)$ is complex. In this notation, the full determinant and the
one for the spatial part are given by
\beq  \mid g \mid =  {1\over 4} \ {\alpha}^2 \ e^{4\phi} , \ \
 \mid g_{ij}\mid =  {1\over 4} \ e^{4\phi}  , \label{det} \eeq
and the line element takes the form

\beq {ds}^2=\alpha^2 {dt}^2 - e^{2\phi} {\mid dz-\beta dt\mid}^2 .
\label{et2} \eeq

The remaining gauge condition will be chosen so as to yield instantaneous
propagation in the equations of motion. This is possible in (2+1) dimensions
because there are not enough transverse coordinates to allow the propagation
of tensor waves, for which a retarded propagator would be needed.

In order to understand better this point it is convenient to rewrite the
Einstein-Hilbert action by splitting the scalar curvature $ R^{(3)} $ into its
space part $R^{(2)}$ and a mixed space-time part as follow \cite{a12}
$(8\pi G=1)$

\beq
S \ = \ - \half\int \sqrt {\mid g\mid } R^{(3)} \ = \ - \half
\int \sqrt {\mid g\mid}
\left[ R^{(2)} +\left( \left(Tr K\right)^2 - Tr K^2 \right)\right] d^3x ,
\label{act} \eeq

where we have dropped a total derivative \cite{a12} giving rise to a boundary
term. In Eq. (\ref{act}) we have introduced the extrinsic curvature tensor
by the expression
\beq
K_{ij} = \sqrt{\frac{|g_{ij}|}{g}} \frac{1}{2} \left( \nabla_i^{(2)} g_{0j} +
\nabla^{(2)}_j g_{0i} - \partial_0 g_{ij} \right) , \ \ \ \ ( i, j = 1, 2 ) ,
\label{ext} \eeq
where we denote by $\nabla_{i}^{(2)}$ covariant derivatives in the space part
of the metric which is also used to raise and lower the space indices ${i}$.

By using the fact that the only nonvanishing component of the 2-dimensional
connection in the gauge (\ref{omp}) is ${\Gamma}^{z}_{zz} = \partial_z
\log {g_{z\bar z}}$ and its complex coniugate, it is easy to realize that the
matrix (\ref{ext}) takes the simple form

\begin{eqnarray}
K_{zz} & = & \frac{1}{2\alpha} e^{2\phi} \partial_z \bar \beta ,
\ \ \ \ K_{\bar z \bar z } = {\bar K}_{zz} , \nonumber \\
K_{z \bar z } & \equiv &  K ( z, \bar z , t ) = \frac{1}{2\alpha}
\left( \partial_z g_{0 \bar z } + \partial_{\bar z} g_{0z} -
\partial_0 g_{z\bar z} \right) = \frac{1}{\alpha}
\Gamma_{0, z \bar z } . \label{red} \end{eqnarray}

Therefore , in this 3-dimensional case, time derivatives only occur in
(\ref{act})
through the expression of $K$ in (\ref{red}). We shall thus set $K=0$, i.e.

\beq \partial_{\bar z} ( {\bar \beta} e^{2\phi} ) +
\partial_z ( \beta e^{2 \phi} ) + \partial_0 ( e^{2 \phi} ) = 0
\label{gau} \eeq

as additional gauge condition.

By using (\ref{omp}) and (\ref{gau}), the action (\ref{act}) becomes simply

\beq S = \int d^3 x \left[ - \alpha \nabla^2 \phi +
\frac{e^{2\phi} }{\alpha} {| \partial_z {\bar \beta} |}^2 \right] .
\label{dac} \eeq

Since the form (\ref{gau}) of the action does not contain time derivatives, it
is now obvious that the propagation of the fields $\alpha, \beta,\phi $ is
instantaneous. As a matter of fact, by adding point-like matter sources,
the Einstein equations derived from (\ref{dac}) are

\begin{eqnarray}
 \nabla^2 \phi  & + & \alpha^{-2} e^{2\phi} \partial_z {\bar \beta}
\partial_{\bar z} \beta = - | g | e^{ - 2 \phi } T^{00} , \nonumber \\
 \nabla^2 \beta  & + & 4 ( 2 \partial_z \phi - \frac{1}{\alpha} \partial_z
\alpha ) \partial_{\bar z} \beta = - 2 |g| e^{-2 \phi}
( T^{0z} - \beta T^{00} ) , \nonumber \\
 \nabla^2 \alpha & - & \frac{2 e^{2\phi}}{\alpha} \partial_z {\bar \beta}
\partial_{\bar z} \beta = \alpha^{-1} |g| ( T^{z\bar z} -
\beta T^{0\bar z} - {\bar \beta} T^{0z} + \beta {\bar \beta}
T^{00} ) , \label{ein} \end{eqnarray}

where $\nabla^2 \equiv 4 \partial_z \partial_{\overline z}$ denotes the
Laplacian,

\beq T^{\mu\nu} = \frac{1}{\sqrt{|g|}} \sum_{(i)} m_i {\left(\frac{dt}{ds_i}
\right)} {\dot\xi}^\mu_i
{\dot\xi}^\nu_i \delta^2 ( {\bf x} - {\bf \xi_i} ( t_i ) ) ,
\ \ \ ( i = 1 , ... , N ) \label{mom} \eeq

is the energy-momentum tensor, $ {{\bf x}} = {\bf \xi_i} (t), v_i, s_i $ are
the
particle trajectories, velocities and proper time, and
${\dot\xi}^\mu_i \ = \  ( 1, v_i )$. It is apparent from (\ref{ein}) that the
fields $\alpha , \beta , \phi $ can now be derived as functions of the
trajectories $\xi_{i} (t)$ and velocities $v_i (t)$ for
any given time. It remains to be checked, however, that the gauge conditions
are consistent with the equations of motion, and that the energy momentum
tensor is conserved.

Let us first remark that, by setting $g_{zz}=g_{\bar z\bar z}=0$ (Eq.
(\ref{omp}))
we have lost the Einstein equation for the corresponding components of the
Ricci tensor $R_{\mu\nu}$, which should therefore be added as constraints,
i.e.,

\beq R_{zz} = T_{zz} , \ \ \ \ R_{\bar z\bar z} = T_{\bar z\bar z} .
\label{ons} \eeq

Furthermore, since the action (\ref{act}) is in general quadratic in the
quantity $K$ given by Eq.(\ref{red}), the additional condition $K=0$ of Eq.
(\ref{gau}) is consistent automatically with the full equations of motion.

It is now not difficult to check (Appendix A) that the constraints (\ref{ons})
and the condition (\ref{gau}) are enough to provide the t-dependence of the
trajectories,
and with proper asymptotic conditions, are indeed equivalent to the covariant
conservation of the energy-momentum tensor, which in turn implies the geodesic
equations

\beq \frac{d^2 \xi_i^\mu}{d {s_i}^2} + {(\Gamma^\mu_{\alpha\beta})}_i
\frac{d\xi^\alpha_i}{ds_i} \frac{d\xi^\beta_i}{ds_i} = 0 , \ \ \ \ \ \ \
( i = 1, .... , N ) \label{geo} \eeq

in the fields provided by Eq. (\ref{ein}).

Therefore our procedure will be to determine first the four fields
$\alpha ,\beta ,\bar\beta ,\phi$ from Eq.(\ref{ein}) in terms of the
trajectories at a given time, and then to determine the trajectories themselves
from the geodesic equations in the self-consistent fields.

This separation of the field equations (\ref{ein}) from the equations of
motion (\ref{geo}) is
essentially due to the 3-dimensional nature of the problem, and is the key
advantage of the conformal-like gauge that we are using. Note that, in
principle, this method allows to find a regular metric and the corresponding
motion for a general set of N moving particles. However, we shall focus
in the following on the perturbative expansion for the two-body system.
%***************************************************************************
%
\section{ Lowest order metric and two-particle motion }
%
%***************************************************************************
The perturbative expansion in $Gm_{i}$ of Eqs. (\ref{ein}) and
(\ref{geo}) is set up
iteratively around Minkowskian metric and linear motion and is rather
straightforward. In fact, by using the espression (\ref{et2}) of the proper
time, we obtain

\beq \frac{dt}{ds_i} = {( \alpha^2 - e^{2\phi} {| v_i - \beta |}^2
)}^{-\frac{1}{2}} |_i  , \label{pro} \eeq

and hence the coefficients of the source terms in the r.h.s. of (\ref{ein})
can all be expressed in terms of $\xi_{i}, v_i$ and of the fields themselves,
evaluated at $\xi_{i}$. As a consequence the (n)-th iteration determines
the source for the (n+1)-th, always through equations of Poisson type.

At first order in $Gm_{i}$, we can use the Minkowskian form of proper time

\beq \frac{dt}{ds_i} = \gamma_i = {( 1 - v_i^2 )}^{-\frac{1}{2}}
\label{min} \eeq

to rewrite Eq. (\ref{ein}) in the linearized form,

\begin{eqnarray}
 \nabla^2 \phi^{(1)} & = & - \sum_{i = 1}^2 \ \gamma_i m_i \delta^{(2)}
( {\bf x} - {\bf \xi_i} ) , \nonumber \\
 \nabla^2 \beta^{(1)} & = & - 2 \sum_{i = 1}^2 \ \gamma_i m_i v_i \delta^{(2)}
( {\bf x} - {\bf \xi_i} ) , \nonumber \\
 \nabla^2 \alpha^{(1)} & = &  \sum_{i = 1}^2 \ \gamma_i m_i v_i
{\bar v}_i \delta^{(2)} ( {\bf x} - {\bf \xi_i} ) .
\label{lin} \end{eqnarray}

Here the inversion of the Laplacian is essentially unique (see later),
and the metric can be solved in terms of the basic fields

\beq \phi_i = - 4 G m_i \log | {\bf x} - {\bf \xi_i} |
\label{bai} \eeq

as follows

\begin{eqnarray}
  \phi^{(1)} & = & \sum_{i = 1}^2 \ \gamma_i \phi_i , \nonumber \\
  \beta^{(1)} & = & \sum_{i = 1}^2 \ 2 \gamma_i v_i \phi_i , \nonumber \\
  \alpha^{(1)} & = & - \sum_{i = 1}^2 \ \gamma_i v_i {\bar v}_i \phi_i .
\label{fir}
\end{eqnarray}

It is a matter of inspection to realize that these solutions solve the
constraints (\ref{ons}) and the gauge condition (\ref{gau}) identically.

The expression (\ref{fir}) is unique, up to the addition of harmonic solutions
of the homogeneous equations in (\ref{lin})
which can be reduced to time-dependent constants by requiring that the
spatial connections $\Gamma^z_{\alpha\beta}$ vanish at space infinity, or in
other terms that rotations are absent at large distances. In principle, this
asymptotic condition still leaves the possibility of adding meromorphic
functions. However, the latter have pole singularities, which would describe
sources with more singular energy-momentum tensors (of, say,
$\delta^{\prime} (z)$
type) which are not considered here.

Finally, one can check that  time-dependent constants cannot be added to
$\phi^{(1)}$
( because of the $ K = 0 $ condition ) nor to
$\beta^{(1)}$ (because of the asymptotic condition) and, as far as
$\alpha^{(1)}$ is concerned, they can be readsorbed in a redefinition of the
time variable. Therefore, we shall take as our starting point the solution
(\ref{fir}), in which the logs are written in units of an arbitrary scale.

To solve for the motion we have to impose the three geodetic equations
(\ref{geo}). First note that the time components can be integrated immediately
using the expression (\ref{et2}) of $ds^2 = g_{\mu\nu}dx^{\mu}dx^{\nu}$, and
yields Eq. (\ref{pro}),
showing that $dt/ds_i$ is given in terms of velocities and fields.

On the other hand, the space components of the geodetic equation have a
simplified structure in
this gauge because the metric at time $t$ only depends on position and
velocities and not on higher time derivatives. As a consequence, since the
affine connection contains only first derivatives of the metric,
Eqs.(\ref{geo})
involve at most the particles' acceleration and are thus of Newtonian type.
Instead in a covariant gauge, due to the retarded propagation,
all time derivatives of $\xi_{i}(t)$ would contribute.

In detail, in order to compute the connection components it is useful to
observe that, in our parametrization, they can be cast in a simple form, which
isolates some first-order contribution. First the $\Gamma^{\mu}_{\nu\rho}$
are expressed in terms of the $\Gamma^{0}_{\nu\rho}$ as follows

\begin{eqnarray}
  \Gamma^z_{zz} & = & \beta \Gamma^{0}_{zz} + 2 \partial_z ( \phi )
\qquad , \qquad \qquad
\Gamma^z_{\bar z\bar z} =  \beta \Gamma^0_{\bar z\bar z} \ ,
\nonumber \\
  \Gamma^z_{0z} & = & \beta \Gamma^0_{0z} - e^{-2\phi} \partial_z
( e^{2\phi} \beta ) \qquad , \quad
\Gamma^z_{0\bar z} =  \beta \Gamma^{0}_{0\bar z} \ , \nonumber \\
\Gamma^z_{00} & = & \beta \Gamma^0_{00} - \partial_0 \beta + e^{-2\phi}
( \partial_{\bar z} \alpha^2 + \beta \partial_z ( e^{2 \phi} \beta )
- e^{2 \phi} {\bar \beta} \partial_{\bar z} \beta ).  \label{gam}
\end{eqnarray}

Furthermore, the $\Gamma^{0}_{\nu\rho}$ themselves are

\begin{eqnarray}
  \Gamma^0_{zz} & = & \frac{1}{2} \alpha^{-2} e^{2\phi} \partial_z
{\bar \beta}  \ , \nonumber \\
  \Gamma^0_{0z} & = & \alpha^{-1} \partial_z \alpha - \beta \Gamma^0_{zz} \ ,
\nonumber \\
  \Gamma^{0}_{00} & = & \alpha^{-1} \partial_0 \alpha - \beta \Gamma^0_{0z}
- {\bar \beta} \Gamma^0_{0\bar z} .
\label{ga2}
\end{eqnarray}

If we limit ourself to the first-order in $G$, all the contributions
proportional to $\beta$ can be neglected, and we arrive at the following
equation:

\beq \frac{d}{dt} \left( \frac{d\xi_1 }{d s_1} \right) =
\frac{d}{dt} \left( \frac{dt}{ds_1} v_1 \right) =
4 G m_2 \gamma_1 \gamma_2 \frac{{( v_1 - v_2 )}^2}{\xi_1 - \xi_2} .
\label{dif} \eeq

To first order accuracy, one can set in the r.h.s. of (\ref{dif})
\beq \dot\xi \equiv {\dot\xi}_1 - {\dot\xi}_2 = v_1 - v_2 =
\frac{P {\cal M}}{E_1 E_2} = V_0 = {\rm const.} , \ \ \ E_i = m_i \gamma_i
\label{not} \eeq
where $ {\cal M} = E_1 + E_2 $. We then obtain
\beq \frac{d}{dt} \left( m_1 \frac{dt}{ds_1} v_1 \right) = g P
\frac{\dot\xi}{\xi} , \ \ \ \ \ ( g \equiv 4 G {\cal M} )
\label{sba}
\eeq
and the constant of motion
\beq P = P_1 = m_1 \frac{dt}{ds_1} v_1 ( 1 - g \log \xi ) . \label{imp}\eeq
By introducing in Eq. (\ref{imp}) the undisturbed trajectory for given impact
parameter $b$
\beq \xi^{(0)} (t) = i b + V_0 t \label{ind} \eeq
we can read off the rotation angle of $v_{1}$ when $t$ varies from $-\infty$ to
$+\infty$, i.e., the first order scattering angle
\beq \theta = \mp \pi g = \mp 4 \pi G {\cal M} , \ \ \ \ ( b {}^{>}_{<} 0 ),
\label{fin} \eeq
a well-known result at this order \cite{a2}, \cite{a4}, \cite{a8}.

By replacing in Eq.(\ref{pro}) the first order fields, we also obtain
\beq {\left( \frac{dt}{ds_1} \right)}^2 [ 1 - {| v_1 |}^2 + 4 G m_2 \gamma_2
{ | v_1 - v_2 | }^2 \log {|\xi|}^2 ] = 1 \label{int} \eeq
and thus by Eq.(\ref{imp}), a constant of motion of energy type
\beq E_1 = m_1 \frac{dt}{ds_1} ( 1 - g \frac{\bar{v}_1 v_2}{2} \log {|\xi|}^2 )
\label{ene} \eeq
such that $E_1^2 - {|P_1|}^2 = m_1^2 $. Therefore, $E_{1}$ and $P_{1}$ have
the meaning of Minkowskian energy and momentum \cite{a8}.

Finally, by using again Eq.(\ref{ene}) to eliminate $ dt / ds_1 $ in
Eq.(\ref{imp}) we obtain a simple expression for the relative speed
\beq \dot\xi (t) = V_0 ( 1 + g \log \xi - g \frac{\bar{v}_1 v_2}{2} \log
{|\xi|}^2 )
\label{spe} \eeq
from which the detailed first order trajectory is easily found.
%***********************************************************************
%
\section{ Second order metric for any speed}
%
%***********************************************************************

At higher orders in $G$, the advantages of working with an instantaneous
gauge show up clearly.
We have already remarked in general that fields and trajectories are determined
by separate equations and that the equations of motion are "Newtonian" in the
sense that they are 2nd order in time at all orders. This allows to find, at
n-th order, the source terms for the (n+1)-th order in a straightforward way.

In practice, at second order in $G_{N}$ we need first to know the first order
correction to proper time. By expanding (\ref{pro}) or (\ref{int}) at first non
trivial order we obtain, say for $i=1$,

\begin{eqnarray}
\frac{dt}{ds_1} & = & \gamma_1 \left[ 1 + \gamma_1^2 \gamma_2 {| v_1 - v_2 |}^2
\phi_2|_1 + ... \right] \nonumber \\
& = & \gamma_1 + \frac{dt^{(1)}}{ds_1} + ...  .
\label{fds} \end{eqnarray}

By replacing (\ref{fds}) and (\ref{fir}) in the r.h.s. of Eqs.
(\ref{ein}) we obtain the field equations in their second order form:

\begin{eqnarray}
\nabla^2 \phi^{(2)} = & - &\ \partial_{\bar z} \beta^{(1)}
\partial_z {\bar \beta}^{(1)} + \left( \alpha^{(1)} \gamma_1 +
{\left( \frac{dt}{ds_1} \right)}^{(1)} \right) \nabla^2 \phi_1  + \nonumber \\
& + & \left( \alpha^{(1)} \gamma_2 + {\left( \frac{dt}{ds_2} \right)}^{(1)}
\right) \nabla^2 \phi_2 , \nonumber \\
\nabla^2 \beta^{(2)} = & + & 4 \partial_z ( \alpha^{(1)} - 2 \phi^{(1)} )
\partial_{\bar z} \beta^{(1)} + \alpha^{(1)} \nabla^2 \beta^{(1)} +
\nonumber \\
& + & 2 \left( {\bar v}_1 {\left( \frac{dt}{ds_1} \right)}^{(1)} - \gamma_1
\beta^{(1)} \right)
\nabla^2 \phi_1 + 2 \left( {\bar v}_2 {\left( \frac{dt}{ds_2}
\right)}^{(1)} - \gamma_2 \beta^{(1)}\right) \nabla^2 \phi_2 , \nonumber \\
\nabla^2 \alpha^{(2)} = & + & 2 \partial_z {\bar \beta}^{(1)}
\partial_{\bar z} \beta^{(1)} -
\left( \gamma_1 v_1 {\bar v}_1 \phi^{(1)} + {\left(\frac{dt}{ds_1}
\right)}^{(1)} v_1
{\bar v}_1 - \gamma_1 ( v_1 {\bar \beta} + {\bar v}_1 \beta )
\right) \nabla^2 \phi_1 - \nonumber \\ & - &
\left( \gamma_2 v_2 {\bar v}_2 \phi^{(1)} + {\left( \frac{dt}{ds_2}
\right)}^{(1)} v_2
{\bar v}_2 - \gamma_2 ( v_2 {\bar \beta} + {\bar v}_2 \beta )
\right) \nabla^2 \phi_2 , \label{sec} \end{eqnarray}

where we have rewritten the $\delta$-functions in terms of Laplacians, and
the first terms in the r.h.s. represent the remaining non-linear parts.

If we carefully read these expressions, using the previous solution for the
fields, it appears that the sources are ill-defined, due to the presence of
self-interactions, like $\phi_{1}\nabla^2\phi_{1}$. However, integrating by
parts
the non linear term produces similar ill-defined terms which exactly cancel
those coming from the sources. At this point the inversion of the  Laplacians
is straightforward, even if cumbersome. After some algebra we get the following
results

\begin{eqnarray}
\phi^{(2)} = & - & \gamma_1^2 v_1 {\bar v}_1 \frac{\phi^2_1}{2} -
\gamma_2^2 v_2 {\bar v}_2 \frac{\phi^2_2}{2} -
\gamma_1 \gamma_2 ( v_1 {\bar v}_2 + {\bar v}_1 v_2 ) \frac{\phi_1
\phi_2}{2} + \nonumber \\
& + & \gamma_1 \gamma_2 ( v_1 {\bar v}_2 - {\bar v}_1 v_2 )
\frac{\phi_{12}}{2} + \nonumber \\
& + & \gamma_1 \gamma_2 \left[ \frac{v_1 {\bar v}_2 + {\bar v}_1 v_2}{2}
- v_1 {\bar v}_1 + \gamma_2^2 ( v_1 - v_2 ) ( {\bar v}_1 -
{\bar v}_2 ) \right] \phi_1|_2 \phi_2 + \nonumber \\
& + & \gamma_1 \gamma_2 \left[ \frac{v_1 {\bar v}_2 + {\bar v}_1 v_2}{2}
- v_2 {\bar v}_2 + \gamma_1^2 ( v_1 - v_2 ) ( {\bar v}_1 -
{\bar v}_2 ) \right] \phi_2|_1 \phi_1  \ , \nonumber \end{eqnarray}

\begin{eqnarray}
\beta^{(2)} = & - & \gamma^2_1 v_1 ( 2 + v_1 {\bar v}_1 ) \phi^2_1
- \gamma^2_2 v_2 ( 2 + v_2 {\bar v}_2 ) \phi^2_2 + \nonumber \\
& - & \gamma_1 \gamma_2 \left[ 2 ( v_1 + v_2 ) + v_1 v_2 ( {\bar v}_1 +
{\bar v}_2 ) \right] \phi_1 \phi_2 + \gamma_1 \gamma_2 \left[ 2 ( v_1 -
v_2 ) - v_1 v_2 ( {\bar v}_1 - {\bar v}_2 ) \right] \phi_{12} +
\nonumber \\
& + & \gamma_1 \gamma_2 \left[ 2 ( v_2 - v_1 ) + v_1 v_2 ( {\bar v}_2 -
{\bar v}_1 ) +
2 \gamma^2_2 v_2 ( v_1 - v_2 ) ( {\bar v}_1 - {\bar v}_2 ) \right]
\phi_1|_2 \phi_2 + \nonumber \\
& + &  \gamma_1 \gamma_2 \left[ 2 ( v_1 - v_2 ) + v_1 v_2 ( {\bar v}_1 -
{\bar v}_2 ) +
2 \gamma^2_1 v_1 ( v_1 - v_2 ) ( {\bar v}_1 - {\bar v}_2 ) \right]
\phi_2|_1 \phi_1 \ , \nonumber \end{eqnarray}

\begin{eqnarray}
\alpha^{(2)}  = & {} & \! \! \! \! \gamma_1^2 v_1 {\bar v}_1 \phi^2_1 +
\gamma_2^2 v_2 {\bar v}_2 \phi^2_2 + \gamma_1 \gamma_2
( v_1 {\bar v}_2 + {\bar v}_1 v_2 ) \phi_1 \phi_2 + \nonumber \\
& + & \gamma_1 \gamma_2 ( {\bar v}_1 v_2 - v_1 {\bar v}_2 )  \phi_{12}
+ \nonumber \\
& + & \gamma_1 \gamma_2 \left[ v_1 {\bar v}_2 + {\bar v}_1 v_2
- 2 v_1 {\bar v}_1 - 2 \gamma_1^2 v_1 {\bar v}_1 ( v_1 - v_2 )
( {\bar v}_1 - {\bar v}_2 ) \right] \phi_2|_1 \phi_1 + \nonumber \\
& + & \gamma_1 \gamma_2 \left[ v_1 {\bar v}_2 + {\bar v}_1 v_2
- 2 v_2 {\bar v}_2 - 2 \gamma_2^2 v_2 {\bar v}_2 ( v_1 - v_2 )
( {\bar v}_1 - {\bar v}_2 ) \right] \phi_1|_2 \phi_2 .
\label{ses} \end{eqnarray}

Here the left over unknown is $\phi _{12}$, which comes from inverting the
Laplacian over the terms with antisymmetric product of derivatives, i.e. is
defined by

\beq \nabla^2 \phi_{12} \ = \ 4 ( \partial_z \phi_1 \partial_{\bar z} \phi_2
- \partial_{\bar z} \phi_1 \partial_z \phi_2 ) .
\label{phi} \eeq

This equation can be integrated directly, but it is more convenient to impose
the gauge condition $K=0$ on the fields (\ref{ses}) from which we get,
after some algebra, the constraint

\begin{eqnarray}
\partial_z \phi_{12} & = & J_z =
( \phi_2 - \phi_2|_1 ) \partial_z \phi_1 - ( \phi_1 - \phi_1|_2 )
\partial_z \phi_2 . \label{inp}
\end{eqnarray}

It is easy to see that (\ref{inp}) yields the Laplacian in Eq. (\ref{phi}), as
expected. Furthermore, Eq. (\ref{inp}) can be used to construct
the solution

\beq \phi_{12} ( z , {\bar z} ) = \int^{(z, {\bar z})}
\ \ ( dz J_{\overline z} \ - \ d{\overline z} J_z ) ,
\ \ \ {\bf J} = ( J_z , J_{\bar z} ) , \label{sol} \eeq

which is automatically single-valued because $\bf J$ is divergenceless,
as a consequence of the subtraction of $\phi_{1}|_{2}$ and
$\phi_{2}|_{1}$ in Eq. (\ref{inp}).

The explicit solution for $\phi_{12}$
(Appendix B) can be written as function of the complex
variable

\beq  Z \equiv \frac{z - \xi_1}{\xi_2 - \xi_1} ,
\label{zed}
\eeq

with $ z = x + i y $ and $ \xi_i = \xi_i^x + i \xi_i^y $, in the form

\begin{eqnarray}
\phi_{12} ( {\bf x},  \xi_1 , \xi_2 ) = & {} & \! \! \! \!
4G^{2}m_{1}m_{2}
\left( - \log ( 1 - Z ) \log {\bar Z} + \log ( 1 - {\bar Z} ) \log Z + \right.
\nonumber \\
& + & \left. Li_2 (1 - Z ) + Li_2 ( {\bar Z} ) - Li_2 ( Z ) - Li_2 (
1 - {\bar Z} ) \right) , \label{log}
\end{eqnarray}

where $Li_{2}(z)$ denotes the Spencer's function \cite{a13}. Using the above
expression we can compute the time derivative of $\phi_{12}$ which can be
written in terms of $\phi_{1}$ and $\phi_{2}$:

\begin{eqnarray}
\partial_0 \phi_{12} = & {} & \! \! \! \! \phi_1 ( v_2 \partial_z - {\bar v}_2
\partial_{\bar z} ) \phi_2 - \phi_2 ( v_1 \partial_z - {\bar v}_1
\partial_{\bar z} ) \phi_1 + \nonumber \\
& - & \phi_1|_2 ( v_2 \partial_z - {\bar v}_2 \partial_{\bar z} ) \phi_2 +
\phi_2|_1 ( v_1 \partial_z - {\bar v}_1 \partial_{\bar z} ) \phi_1 +
\nonumber \\
& + & \left( ( v_1 - v_2 ) (\partial_z \phi_2)|_1 - ( {\bar v}_1 - {\bar v}_2 )
( \partial_{\bar z} \phi_2 )|_1 ) \right) \phi_1 +
\nonumber \\
& + & \left( ( v_1 - v_2 ) (\partial_z \phi_1)|_2 - ( {\bar v}_1 - {\bar v}_2 )
( \partial_{\bar z} \phi_1 )|_2 ) \right) \phi_2 .
\label{dph} \end{eqnarray}

These relations are useful to check that the gauge condition (\ref{gau}) is
satisfied, once the first order geodetic motion is taken into account.
%***************************************************************************
%
\section{ Equations of motion and scattering angle}
%
%***************************************************************************
Studying the equations of motion (\ref{geo}) involves replacing the cumbersome
second order fields (\ref{ses}) in the expressions (\ref{gam}) and
(\ref{ga2}) for the affine
connection, and gives rise to a rather lengthy algebra. The latter is however
simplified by the following observations:

{1)} All singular terms containing at least one field $\phi_{i}$,
evaluated at the source, should cancel out. In other words, there are no
self-interactions.

{2)} In the r.h.s. of (\ref{geo}) one can use the first order equations of
motion, which involve several conserved quantities, described in Eqs.
(\ref{imp}) and (\ref{ene}). In particular one can define a c.m. frame in which

\beq m_1 \frac{dt}{ds_1} v_1 + m_2 \frac{dt}{ds_2} v_2 = 0 ,
\label{bar} \eeq

at least up to second order in $Gm_{i}$.

More precisely, after doing the algebra mentioned before, we arrive at the
following equation for , say, the spatial components of particle 1

\begin{eqnarray}
 \frac{d}{ds_1} \left( \frac{dt}{ds_1} v_1 \right) = & {} & \! \! \! \!
{( \frac{dt}{ds_1} )}^2 \left[ 2 \gamma_2 { ( v_1 - v_2 )}^2 (\partial_z
\phi_2)|_1 ( 1 + \gamma_1 \gamma_2^2 {| v_1 - v_2 |}^2 \phi_1|_2 -
\gamma_1 v_1 ( {\bar v}_1 - {\bar v}_2 ) \phi_1|_2 ) \right. + \nonumber \\
& + & \left. 2 \gamma_2 v_2 (\partial_{{\bar z}} \phi_2 )|_1 {( {\bar v}_1 -
{\bar v}_2 )}^2 ( \gamma_2 v_2 \phi_2|_1 + \gamma_1 v_1 \phi_1|_2 ) \right] ,
\label{ege} \end{eqnarray}

while the time component can be replaced by the expressions of $dt/ds_{1}$ and
$dt/ds_{2}$ obtained from Eq. (3.1).

To second order accuracy one can use the first order equations of motion in
the r.h.s. of (\ref{ege}), and in particular the expression
(\ref{fds}) for $dt/ds_{2}$
and the center of mass frame condition (\ref{bar}), to obtain
\begin{eqnarray}
\frac{d}{dt} \left( m_1 \frac{dt}{ds_1} v_1 \right) = 4 G m_1 m_2
\frac{dt}{ds_1} \frac{dt}{ds_2} \frac{{(v_1 - v_2 )}^2}{\xi_1 - \xi_2}
( 1 - \gamma_1 v_1 ( {\bar v}_1 - {\bar v}_2 ) \phi_1|_2 ) .
\label{sim}\end{eqnarray}

The discussion of this equation can be further simplified by introducing the
Minkowskian energies $E_{1}$ and $E_{2}$ and momentum $P$, which appear as
first-order constants of motion in Eqs. (\ref{imp}) and (\ref{ene}). By using
the notation

\beq V_1 = \frac{P}{E_1} , \ \ \ V_2 = - \frac{P}{E_2} , \ \ \
{\cal M} = E_1 + E_2 , \ \ \ g = 4 G {\cal M} , \ \xi = \xi_1 - \xi_2
\label{foc} \eeq

Eq. (\ref{sim}) can be rewritten as

\beq \frac{d}{dt} \left( m_1 \frac{dt}{ds_1} v_1 \right) = 4 G E_1 E_2
\frac{{\dot\xi}^2}{\xi} \left( 1 + \frac{g}{2} {\bar V}_1 V_2 \log {| \xi |}^2
\right) , \label{ide} \eeq

where, by the first order equation (\ref{int}), we can set

\beq {\dot\xi} = ( V_1 - V_2 ) ( 1 + g \log \xi - \frac{g}{2} {\bar V}_1 V_2
\log {|\xi|}^2 ) + O ( G^2 ) . \label{fov} \eeq

By finally replacing (\ref{fov}) in the r.h.s. of (\ref{ide}) we obtain

\beq \frac{d}{dt} \left( m_1 \frac{dt}{ds_1} v_1 \right) = g P \frac{\dot\xi}
{\xi} ( 1 + g \log \xi ) , \label{exp} \eeq

which can be integrated to yield

\beq p_1 ( t ) \equiv m_1 \frac{dt}{ds_1} v_1 = P ( 1 + g \log \xi +
\frac{1}{2} g^2 {( \log \xi )}^2 )  . \label{fif} \eeq

In conclusion, the "momentum" \ variable ${p}_{1}(t)$, as \ function of
\ the relative distance $\xi$,
just exponentiates the first order result and, up to
second order included, has by (\ref{fov}) and (\ref{fif}) the form

\beq p_1 ( t ) = P \xi^{g} \simeq P {(V_0 t + i b )}^{g(1+g)}
{| V_0 t |}^{-g^2 {\bar v}_1 v_2} ,  \ \ \ \
( |V_0 t | \gg b ) . \label{asy} \eeq

{}From the large time behaviour in (\ref{asy}) we can read off the second order
scattering angle

\beq \theta = \mp \pi g ( 1 + g ) + O (g^3) , \ \ \ \ g \equiv 4 G {\cal M} ,
\ ( b > 0 ) \ \ ( ( b < 0 ) ), \label{obv} \eeq

an expression which can be checked by explicit integration of Eq.
(\ref{asy} ) to
yield $\xi_{1}(t)$ at all times.

The results (\ref{asy}) and (\ref{obv}) call for several comments.
Firstly, the impressive
simplification of nonlinearities in this gauge is presumably rooted in a
simple relation to the Minkowskian (singular) gauge \cite{a2}, \cite{a8} which
may hold in this case.
In fact the present instantaneous gauge is actually equivalent to a
Coulomb-type gauge \cite{a5}, \cite{fer}
in a first order (Palatini) formalism and this may provide a basis for a
non-perturbative construction \cite{a14} of dreibein and metric to all orders.
Secondly, the expression (\ref{obv}) shows no explicit $m_{i}$ dependence at
fixed
total invariant mass ${\cal M}$, which in turn coincides with the topological
invariant \cite{a4} at this order. Thus, there  is a smooth massless limit
and there are second order corrections to the scattering angle even in the
massless case. This is in agreement with suggestions by 't Hooft
\cite{a5} , and is at variance with previous findings by Cappelli, and two
of us \cite{a8} in covariant-type gauges, which provide an alternative
definition
of c.m. frame, \cite{a15}.

Although disappointing, the gauge dependence of the scattering angle noticed
above is not terribly surprising because the instantaneous gauge changes
in a profound way the relation of two-body vs. one-body metrics:  in
particular there is no simple way of decoupling particles at large space
separations
due to the presence of logaritmically increasing fields. This is  to be
contrasted to what happens in covariant-type gauges \cite{a8}, \cite{a15} where
such decoupling is built in and gives rise, in the massless limit, to
scattering of Aichelburg-Sexl type.

The above remarks show that further study of our conformal type gauge is
needed,
possibly at non perturbative level, in order to better investigate the role of
asymptotic conditions in the scattering problem.

{\bf Acknowledgements}

It is a pleasure to thank Stanley Deser and Giorgio Longhi for interesting
discussions and suggestions.

%***************************************************************************
\appendix
\section{ - Constraints vs. equations of motion}
%
%***************************************************************************

In order to discuss the consistency of the instantaneous gauge with the
geodetic motion, it is useful to recast the field equations
(\ref{ein}) in terms of new variables  $\gamma_{\mu\nu} = h_{\mu\nu} - \half
h_{\alpha}^{\alpha} \eta_{\mu\nu}$ , where $h_{\mu\nu} = g_{\mu\nu} -
\eta_{\mu\nu}$
with the property  $\gamma_{12} = 0 , \gamma_{11} =
\gamma_{22}$ . These can be
rewritten as four basic equations
\begin{eqnarray}
\half  \nabla^2 ( \gamma_{00} - \gamma_{11} ) & = & \Lambda_{00} + g
T_{00} \ = \  \tilde{T}_{00} \nonumber \\
\half  \nabla^2  \gamma_{0i} & = & \Lambda_{0i} +  g T_{0i} \ = \
\tilde{T}_{0i} \nonumber \\
\nabla^2  \gamma_{11} & = & \Lambda_{xx} + \Lambda_{yy} + g ( T_{xx} +
T_{yy} ) =  ( \tilde{T}_{xx} + \tilde{T}_{yy} ) \label{bas} \end{eqnarray}
where the tensor $\Lambda_{\mu\nu}$ is given by
\beq \Lambda_{\mu\nu} \ = \ \frac{1}{4} \ \eta_{\mu\mu'} \ \eta_{\nu\nu'} \
\epsilon^{\mu'\rho\sigma} \ \epsilon^{\nu'\gamma\delta} \ g_{\alpha\beta} \
\left[ \ \Gamma^{\alpha}_{\sigma\gamma} \Gamma^{\beta}_{\rho\delta} -
\Gamma^{\alpha}_{\sigma\delta} \Gamma^{\beta}_{\rho\gamma} \ \right] , \eeq
and the modified energy-momentum tensor $\tilde{T}_{\mu\nu}$ \ satisfies
the \ trivial \  conservation law
$\partial_\mu \tilde{T}^{\mu\nu} = 0$
equivalent to the covariant conservation of $T^{\mu\nu} $.

The other two Einstein equations give constraints on the integration of the
four variables $\gamma_{\mu\nu}$. These can be summarized in one complex
equation
\beq G_{zz} = \partial_z ( \partial_0 \gamma_{0z} - \partial_{z} \gamma_{11}) -
\Lambda_{zz} -  g T_{zz} = 0 . \label{kcm} \eeq

The gauge condition $ K = 0 $ can also be rewritten as

\beq \partial_i \gamma_{0i} = \partial_0 ( \gamma_{00} - \gamma_{11} )
\label{k00} \eeq

A consistency test is provided by imposing that the Laplacian of the
gauge condition (\ref{k00}) and of the complex equation (\ref{kcm}) is zero.
Using the first four equations of motion (\ref{bas}) we get
\beq \half \nabla^2 [ \partial_0 ( \gamma_{00} - \gamma_{11} ) -
\partial_i ( \gamma_{0i} ) ] = \partial_{\alpha} \tilde{T}_{0\alpha} = 0
\label{emp} \eeq
\beq \nabla^2 ( G_{zz} ) = \partial_z \partial_{\alpha} \tilde{T}_{z\alpha} = 0
\label{str} \eeq
Hence (\ref{emp}) and (\ref{str}) show that, for every solution of the first
four  equations,
imposing the covariant conservation of $T^{\mu\nu}$, equivalent to the
geodesic equations (\ref{geo})implies that the gauge condition and the
constraint on $G_{zz}$ are simply the sum of pure analytic and anti-analytic
functions. Requiring that the connections vanish at spatial infinity,
i.e. imposing that $K = 0$ and $G_{zz} = 0$ as a boundary condition, is then
enough to ensure that these equations are satisfied in the whole
two-dimensional
plane, as stated in Sec. II.

%****************************************************************************
%
\section{ - Monodromic solution for Poisson-like equation }
%
%****************************************************************************
In the following we show how to construct a single-valued solution to
the Poisson-like equation (\ref{phi})in two spatial dimensions avoiding the
nasty calculations implied by the more general Green-function method.

Since $\nabla^{2} = 4\partial_{z}\partial_{\bar z}$, it is easy
to obtain a particular solution of such type of equations by
just integrating in $z$ and $\bar z$ the source, which is
given as a sum of terms with factorized dependence on $z$ and $\bar z$.

The integration may produce unwanted polydromy. Since the source is
monodromic, then the polydromic terms have simple discontinuities which are
analytic or anti-analytic functions and can be eliminated by exploiting
the arbitrariness in the inversion of the  Laplacian \cite{acv}.

In our case, by integrating eq. (\ref{phi}) we obtain the following particular
solution
\beq (4G^{2}m_{1}m_{2})^{-1} \ \phi_{12}^0 =
\log ( z - \xi_1 )
\log ( {\bar z } - {\bar \xi_2} ) - ( 1 \leftrightarrow 2 )  \label{par2} \eeq
If we circle particle $1$, then the r.h.s. of (B.1)
gets an additional contribution
\beq + \  2 \pi i  [ \log ( {\bar z } - {\bar \xi_2} ) - \log ( z - \xi_2 ) ] .
\eeq

To compensate for the previous contribution we need to add the following
harmonic function which has the opposite discontinuity
around particle $1$ of the particular solution and no discontinuity around
particle $2$:
\beq
h_1= Li_2 ( 1 - Z ) - \log ( Z ) \log ( {\bar\xi}_1 - {\bar\xi}_2 ) - c.c.  ,\\
\eeq
where the Spencer function $Li_{2}(z)$ has a branch-point at $z=1$ and is
defined by
\beq Li_{2}(z)= -\int_{0}^{z} \frac {dx}{x} \ln (1-x) = \sum_{n=1}^{\infty}
\frac{z^{n}}{n^2}  ,
\eeq
and we remind that $Z= (z-\xi_{1})/(\xi_{2} - \xi_{1})$. Similarly
to compensate for the discontinuity of $\phi^0_{12}$ around the particle
$2$, we need to add an other harmonic function
\beq
h_2= - Li_2 ( Z ) + \log ( 1 - Z ) \log ( {\bar\xi}_2 - {\bar\xi}_1 ) - c.c.
.\\
\eeq
By adding (B.3) and (B.5) to the r.h.s. of (B.1), we get
the complete single-valued solution $\phi_{12}$ given in Eq. (\ref{log})
of the text.

%****************************************************************************
%

\end{document}